\RequirePackage{cmap} 
\documentclass[aps,pra,showpacs,twoside,10pt,floatfix,nofootinbib,longbibliography,twocolumn]{revtex4-2}

\usepackage{lmodern}
\usepackage[T1]{fontenc}
\usepackage[utf8]{inputenc}

\usepackage{geometry}
\usepackage{amssymb,amsfonts,amsmath,bm,amsthm,mathtools,braket}

\usepackage{newtxtext}
\usepackage{newtxmath}

\usepackage{xcolor,comment,wrapfig}
\usepackage[normalem]{ulem}
\usepackage[colorlinks=true, citecolor=red, urlcolor=blue]{hyperref}

\usepackage[most]{tcolorbox}

\definecolor{checkcolor}{HTML}{7541C0}

\definecolor{boldcolor}{HTML}{7541C0}
\definecolor{boldcolor1}{HTML}{4A2166}
\definecolor{boldcolor2}{HTML}{6C3391}
\definecolor{boldcolor3}{HTML}{366E8A}
\definecolor{boldcolor4}{HTML}{69B2D6}

\newcommand{\THEO}[1]{\vspace{0.1cm}\noindent\textbf{\textcolor{boldcolor1}{$\blacksquare$ Theorem #1.}}}

\newcommand{\PROP}[1]{\vspace{0.1cm}\noindent\textbf{\textcolor{boldcolor2}{$\blacksquare$ Proposition #1.}}}

\newcommand{\CORL}[1]{\vspace{0.1cm}\noindent\textbf{\textcolor{boldcolor3}{$\Diamond$ Corollary #1.}}}

\begin{document}

\title{
Quantum advantage unlocked: Charging quantum batteries with \texorpdfstring{$K$}{K}-regular graph stabilizers
}
\author{Debkanta Ghosh$^{1,2}$,  Tanoy Kanti Konar$^{1,2,3}$, Gianluca Francica$^4$,  Amit Kumar Pal$^5$, Aditi Sen(De)$^{1,2}$}
\affiliation{$^1$Harish-Chandra Research Institute, Chhatnag Road, Jhunsi, Allahabad - 211019, India\\
$^2$Homi Bhabha National Institute, Training School Complex, Anushakti Nagar, Mumbai 400 094, India\\
$^{3}$Mark Kac Complex Systems Research Center, Jagiellonian University in Krak\'ow, PL-30-348 Krak\'ow, Poland\\
$^4$Dipartimento di Fisica e Astronomia, Università di Padova, via Marzolo 8, 35131 Padova, Italy  \\
$^5$Department of Physics, Indian Institute of Technology Palakkad, Palakkad 678 623, India}

\begin{abstract}
Regular graphs find broad applications ranging from quantum communication to quantum computation. Motivated by this, we investigate the design of a quantum battery based on a  \(K\)-regular graph, where \(K\) denotes the number of edges incident on each vertex. We show that a \(0\)-regular graph battery exhibits extractable work that scales linearly with the system-size when charged using a \(K\)-regular graph. This linear scaling is shown to persist even when the charging is implemented via a collective \(K\)-regular charger with power-law decaying interactions. Interestingly, we prove that both the maximum average power and instantaneous power scale super-linearly in the thermodynamic limit when the connectivity of the charging graph is of the order of the system size, thereby exhibiting {\it quantum advantage}. Furthermore, by introducing the notion of the fraction of extractable work when only subsystems are accessible, we identify this fraction  to be  independent of system-size if the battery is prepared in the down-polarized product state.  This independence breaks down when the battery is oriented along the \(x\)- and \(y\)-directions of the Bloch sphere.

\end{abstract}

\maketitle

\section{Introduction}
\label{sec:intro}

Energy storage devices play a crucial role across a wide range of technologies, from communication systems to medical equipments. In this context, the pioneering work of Alicki and Fannes \cite{Alicki} demonstrated that ensembles of $N$  non-interacting 
$d$-level quantum systems can exploit collective quantum effects to enhance both energy storage and extraction \cite{Binder2015,campaioli2017,Giovannetti2019,barra2019}. Since then, numerous quantum battery (QB) models have been proposed~\cite{andolina2019,santos2019,srijon2020,dou2022LMGQB,le2018,andolina2017,santos2020,cres2020,dou2022,Hu_2022,rossini2020,konar_battery_1,abah2022,srijon21,liu2019,quach2020,tejero2021,Hovhannisyan2013,gyhm2022,Gyhm2024,Francica2024,grazi2025}, where the battery is initialized in an eigenstate of either a non-interacting, or an interacting quantum spin model, and subsequently charged via global or local unitary operations, thereby highlighting the role of correlations among its subsystems~\cite{Binder2015}. Along with putting forward the proposals of several QB models \cite{Batteryreview,campaioli2018quantumbatteriesreview,battery_rmp_review}, extensive efforts have been given in optimizing QB performance by combining insights from quantum information theory with many-body physics~\cite{andolina2017,andolina2020,Bera2020QB,ksen_battery_1,Mitchison2021chargingquantum,alba_1_20,cres2020,konar_battery_2,ksen_battery_4,Francica2022,arjmandi_pra_2022,santos_pra_2023,mazzoncini_pra_2023,ksen_battery_5,AI_quantum_battery,sashi_pra_1,remote_charging_battery,NV_battery,sashi_2,ahmdi_prl_2024,grazi_prl_2024,downing_pra_2024,rinaldi_pra_2025,yadav2025,sarkarpc2025,cavaliere2025}. These works aim to minimize charging time~\cite{andolina_prl_2025}, maximize extractable energy~\cite{perciavalle2024}, and enhance robustness against decoherence \cite{Giovannetti2019,srijon21}. Advanced strategies, including the utilization of topological properties~\cite{topological_quantumbattery},  non-Hermitian framework \cite{konar_battery_3,ahmadi_prl_2025,yang_pra_2025,zhou2025}, indefinite causal order~\cite{battery_ico,simonov_pra_2022,li_pra_2025}, daemonic ergotropy~\cite{Francica2017}, many-body localization~\cite{rossini_prb_2019,arjmandi_pre_2023}, measurement-based scheme~\cite{mesure_battery}, time-crystal~\cite{sahoo2025} and charging multi-mode batteries via a single auxiliary~\cite{pushpan2026}, have also been employed to boost battery's performance. Many of these theoretical predictions have already been experimentally validated in various platforms, such as quantum dots~\cite{wennigerexpqdots}, superconducting circuits~\cite{superconducting_battrey_1,superconductQBexp,GemmeexpIBMsupercond}, organic semiconductors~\cite{Quach2022Jan}, and nuclear magnetic resonance~\cite{MaheshexpNMR}.

Despite extensive theoretical and experimental progress, the deeper potential of quantum batteries  for quantum information processing tasks is only partially understood. A recent development~\cite{kurman2025QC_QB} has shown that QBs can operate as intrinsic energy sources for gate-based quantum computation, providing a new direction for integrating energetic resources directly into computational architectures.
Motivated by this perspective, we examine the performance and thermodynamic behavior of QBs when the charging mechanism is governed by graph-based stabilizer Hamiltonians~\cite{nielsenchuang2000,hein2006entanglement}. In particular, a graph \(G\) is defined through a vertex set (\(V(G)\)), an edge set (\(E(G)\)), and a rule that assigns each edge to a pair of vertices it connects. Precisely, a 
\(K\)-regular graph state is an  entangled stabilizer state associated with a graph having a \emph{regularity} $K$, i.e.,  every vertex in the graph has exactly \(K\) neighbors.  Hence, the symmetry and entanglement structure \cite{hein_pra_2004} inherent to these states have the potential to play a crucial role across a broad spectrum of quantum technologies~\cite{Podzie2025}, including quantum computation~\cite{briegel2009measurement}, quantum error correction~\cite{hu_pra_2008,looi_pra_2008}, and quantum communication protocols such as quantum repeaters~\cite{li_prl_2025} and quantum secret sharing~\cite{markham_pra_2008}. This work seeks to understand how 
the graph connectivity, quantified by the number of neighbors \(K\), influences the system’s capacity to store and deliver energy, and identify the optimal connectivity that maximizes the extractable work from the battery encoded within the lattice, leading eventually to a quantum advantage.

\begin{figure*}
    \centering
    \includegraphics[width=0.8\linewidth]{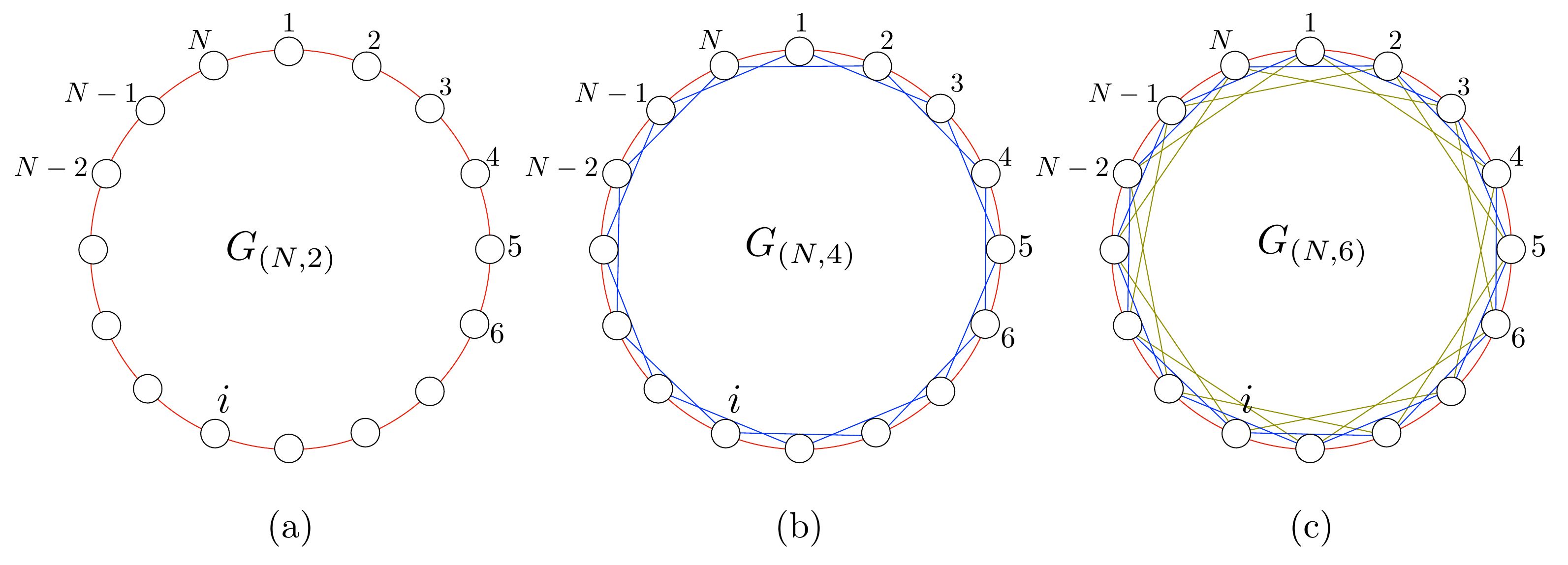}
    \caption{(Color online.) In a $K$-regular ($K$ even) connected graph $G_{(N,K)}$ of $N$ nodes, each node, indexed by $i$ ($i=1,2,\cdots,N$), is connected to $K$ nodes such that the links $\{(i,i+j),(i,i-j);j=1,2,\cdots,K/2\}\in G_{(N,K)}$. Examples with (a) $K=2$, (b) $K=4$, and (c) $K=6$ are shown for arbitrary $N$. Note that each $G_{(N,K)}$ corresponds to a stabilizer Hamiltonian $H_{(N,K)}$ (see Eq.~(\ref{eq:HNK})), while all three graphs shown in (a)-(c) contribute to constitute $H_C=\sum_{K=2}^6J_K H_{(N,K)}$, with $6\leq K_{\max}=N-2$ $(N-1)$ for even (odd) $N$.}
    \label{fig:GNK}
\end{figure*}

Towards addressing these issues, we first establish that the amount of extractable work is invariant under an exchange of the battery and charger Hamiltonians when both are constituted of generators of $K$-regular graphs with different $K$. We examine two distinct scenarios. (a) In one, the charger is designed with the graph generators of a specific $K$-regular graph, while (b) in the other, the charging Hamiltonian involves the graph generators of a collection of graphs with different regularities, where the strength of the interaction corresponding to different values of $K$ decays as a power law. In both scenarios, the battery is \emph{local}, i.e., modeled as an ensemble of $N$ non-interacting spin-$1/2$  particles oriented along $x$, $y$, or $z$ directions, which is the special case of $K=0$. It is well known that a superlinear scaling of extractable work with system-size signals a {\it quantum advantage} \cite{Bera2020QB}. Within this framework, we establish that in scenario (a), super-linear scaling can be obtained at the thermodynamic limit, i.e., \(N\to\infty\) when the battery is being charged via a \(K\)-regular graph with \(K\sim N\). Specifically, we prove that both instantaneous and maximum average power scale as \(N^{3/2}\) when both \(N\) and \(K\) are large and $K\sim N$, depicting the role of connectivity in obtaining quantum advantage in battery charging. On the other hand, such quantum advantage diminishes when \(K\) is kept fixed, and a linear scaling of power is observed. This linear scaling explicitly demonstrates how collective quantum effects can still be exploited in a controlled manner, albeit without yielding superlinear enhancement.

In the scenario (a), our analysis further  reveals that the  extractable work remains independent of the regularity, $K$, of the graph provided the initial local battery Hamiltonian is aligned along the $z$-axis. In contrast, for the initial battery Hamiltonian oriented along the $x$- or $y$-directions, the work output increases with increasing $K$ of the graph involved in the charging process. We additionally consider scenarios in which  only a subset of the full system is accessible for work extraction. In this case, the fraction of extractable energy remains independent of \(N\)  for the initial battery Hamiltonian oriented along the $z$-directions, while such independence is lost for initial battery alignments along the $x$- or $y$-directions. Moreover, in the scenario (b) involving $K$-dependent interactions among graph generators with power-law decay, we find that the average power continues to scale linearly with system-size, confirming the absence of any system-size–dependent gain in the $K$-regular graph-based quantum batteries.

The rest of this paper is organized as follows: In Sec.~\ref{sec:stabilizer}, we introduce the set-up and discuss the connection between the $K$-regular stabilizer batteries and chargers. We also introduce the definitions of stored work, power,  and ergotropy used throughout the paper.   In Sec.~\ref{sec:K_uniform_charger}, we discuss the charging of local batteries using chargers based on $K$-regular graph stabilizers, and demonstrate the existence of quantum advantage in the thermodynamic limit with $K\sim N$. In Sec. \ref{sec:fracen}, we define  the fraction of extractable energy  and its behavior including saturation with system-size $N$. Sec.~\ref{sec:collective} analyzes charging with a charging Hamiltonian constituted of generators corresponding to different $K$-regular graphs with power-law interaction strength. Concluding remarks are given in Sec.~\ref{sec:conclusion}.

\section{Designing \texorpdfstring{$K$}{K}-regular  batteries and chargers}
\label{sec:stabilizer} 

The working principle of a quantum battery involves two key components, namely, (i) the battery Hamiltonian $H_B$ which typically fixes the initial battery-state as its ground state $\ket{\psi_B}$, i.e., $\rho(0)=\ket{\psi_B}\bra{\psi_B}$ at $t=0$,  and (ii) the charging Hamiltonian $H_C$, which, when turned on at $t>0$, governs the energy storage process. We confine ourselves to $H_B$ and $H_C$ having the form 
\begin{eqnarray}
    \label{eq:HNK}
    H_{(N,K)}= \sum_{i=1}^NH_{(i,K)},
\end{eqnarray}
with 
\begin{eqnarray}
H_{(i,K)}&=&\left[\bigotimes_{j=1}^{K/2}Z_{i-j}\right]X_i\left[\bigotimes_{j=1}^{K/2}Z_{i+j}\right],
    \label{eqn:weigh_stab}
\end{eqnarray}
where $K$ is even, and has a maximum value $K_{\max}=N-2$ ($N-1$) for even (odd) $N$ with $N>2$. Note that each $H_{(i,K)}$ for a fixed $K$ can be interpreted as the stabilizer graph generator~\cite{hein2006entanglement} corresponding to the node $i$ in a $K$-regular connected graph $G_{(N,K)}$ of $N$ nodes (see Fig.~\ref{fig:GNK}) with $K$ even. Each node $i$ in $G_{(N,K)}$ is connected to $K$ nodes in such a way that the links $\{(i,i+j),(i,i-j)\}$, $j=1,2,\cdots,K/2$ exists in $G_{(N,K)}$,  the operators at site \(i\),  $A\in\{X,Y,Z\}$ with  $X, Y$, and $Z$ being the Pauli matrices and periodic boundary condition is assumed, i.e., $A_{N+i}\equiv A_i$. This allows interpreting $H_{(N,K)}$  as the  stabilizer Hamiltonian corresponding to the graph $G_{(N,K)}$ constituted of the graph generators. For brevity, in the following, we refer to $H_{(N,K)}$ as the \emph{$K$-regular Hamiltonian}, and the battery (charger) described by this Hamiltonian as the \emph{$K$-regular battery} (\emph{$K$-regular charger}).  Note that by virtue of the properties of stabilizer graph generators corresponding to a graph with fixed $K$-regularity,
\begin{eqnarray}
    \left[H_{(i,K)},H_{(i^\prime,K)}\right]=\left[H_{(N,K)},H_{(i,K)}\right]=0,
\end{eqnarray}
while 
\begin{eqnarray}
    \left[H_{(N,K)},H_{(N,K^\prime)}\right]\neq 0.
    \label{eq:non_commutation}
\end{eqnarray}

\begin{figure}
    \centering
    \includegraphics[width=0.8\linewidth]{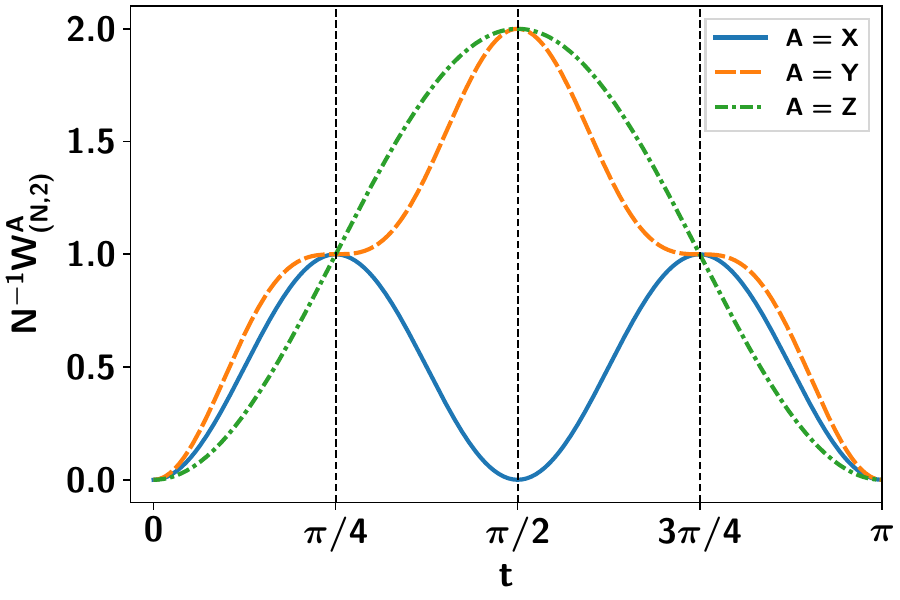}
    \caption{(Color online.) Variations of work stored over the system-size, $N^{-1}W^A_{(N,2)}$ (vertical axis) with time, $t$ (horizontal axis) for the charger, $H_{(N,2)}$. Solid, dashed and dashed-dot lines signify the orientation of the initial battery Hamiltonian, $X,Y$, and $Z$ respectively. Note that the time period of $N^{-1}W^A_{(N,2)}$ is  $t=\pi$ for $A=Y$ and $Z$ with the maximum value $2$ occur at $t=\pi/2$ whereas $(N^{-1}W^X_{(N,2)})_{\max}=1$ at $t=\pi/4$. All the axes are dimensionless.}
    \label{fig:avg_work}
\end{figure}


\subsection{Figures of merit}

We consider a $K^\prime$-regular battery to be prepared in an isolated fashion so that the time-evolution of $\rho(0)$ generated by turning on the $K$-regular charger ($K\neq K^\prime$) is unitary, leading to 
\begin{eqnarray}
    \rho(t)=U\rho(0)U^\dagger, \text{with } U=\exp\left[-\text{i}H_C t\right],
\end{eqnarray}
with $H_C=H_{(N,K)}$ as the charging Hamiltonian. The performance of the $K^\prime$-regular battery  can be quantified through the \emph{work output}, given by 
\begin{eqnarray}
\label{eq:maximum_extractable_work}
    W_{(N,K^\prime,K)} (t) = \text{Tr}\left[\{\rho(t) - \rho(0)\} H_B\right],
\end{eqnarray}
with $H_B=H_{(N,K^\prime)}$, where we have introduced the subscript $(N,K^\prime,K)$ to keep track of the battery and the charging Hamiltonian.  Since the evolution is unitary, the work coincides with the \emph{maximum extractable work} when the initial state is the ground state of the battery~\cite{Alicki}, referred to as the \emph{ergotropy}~\cite{Allahverdyan2004}, and defined as 
\begin{eqnarray}
    \mathcal{E}_{(N,K^\prime,K)}(t)&=&\text{Tr}[\rho(t) H_B] - \underset{U}{\min}\text{Tr}[U\rho(t)U^\dagger H_B],
    \label{eq:ergo}
\end{eqnarray}
where the minimization is performed over all unitary operators. The \emph{maximum average power} of the battery, quantifying the speed of charging during an interval $[0,T]$,  is defined as
\begin{eqnarray}
    \overline{P}_{(N,K^\prime,K)}&=&\max_{t} t^{-1}W_{(N,K^\prime,K)}(t),\;t\in[0,T],
    \label{eqn:avg_power}
\end{eqnarray}
while the \emph{instantaneous power} is given by\cite{battery_rmp_review} 
\begin{eqnarray}
    P_{(N,K^\prime, K)} &=& \frac{d}{dt}[{W}_{(N,K^\prime,K)}(t)].
    \label{eqn:inst_power}
\end{eqnarray}
%

\subsection{Interchangeable batteries and chargers} 

Note that $H_{(N,K)}$s corresponding to all $K$ can be obtained via   
a suitable unitary transformation $\mathcal U_K$ on $H_{(N,0)}$ as 
\begin{equation}
    H_{(N,K)} = \mathcal{U}_K\, H_{(N,0)}\mathcal{U}_K^\dagger,
\end{equation}
with 
\begin{equation}
    \mathcal{U}_{K}=\prod_{j=2,4,\ldots, K} \left(\prod_{i=1}^{N} \mathrm{CZ}_{i,i+j/2}\right),
\end{equation}
where $\mathrm{CZ}_{i,i+j/2}$ is a controlled-$Z$ gate acting on sites $i$ and $i+j/2$ leading to $\mathcal{U}_K=\mathcal{U}_K^\dagger$,  allowing one to efficiently simulate $H_{(N,K)}$ in a circuit model. Moreover, $H_{(N,K^\prime)}$ and $H_{(N,K)}$ (with $K^\prime < K$) can be connected via a unitary operator $\mathcal{U}_{(K^\prime,K)}$ as 
\begin{eqnarray}
    H_{(N,K)}=\mathcal{U}_{(K^\prime,K)} H_{(N,K^\prime)} \mathcal{U}_{(K^\prime,K)}^\dagger, \label{eq:stab-to-stab-unitary}
\end{eqnarray}
where 
\begin{equation}
    \mathcal{U}_{(K^\prime,K)}=\prod_{j=K^\prime+2, K^\prime+4, \ldots, K } \left(\prod_{i=1}^{N} \mathrm{CZ}_{i,i+j/2} \right),
\end{equation}
with $\mathcal{U}_{(K^\prime,K)}=\mathcal{U}^\dagger_{(K^\prime,K)}$. Consequently,  roles of the battery and charging Hamiltonians can be interchanged without affecting the amount of work stored in the battery. This is formalized as the following proposition (see Appendix~\ref{app:proof_prop_1} for the proof).

\PROP{1} \emph{For $H_B$ and $H_C$ both belonging to the class of $K$-regular Hamiltonians, interchanging $H_B$ and $H_C$ keeps the stored work unchanged.}

\begin{figure*}
    \centering
    \includegraphics[width=0.7\linewidth]{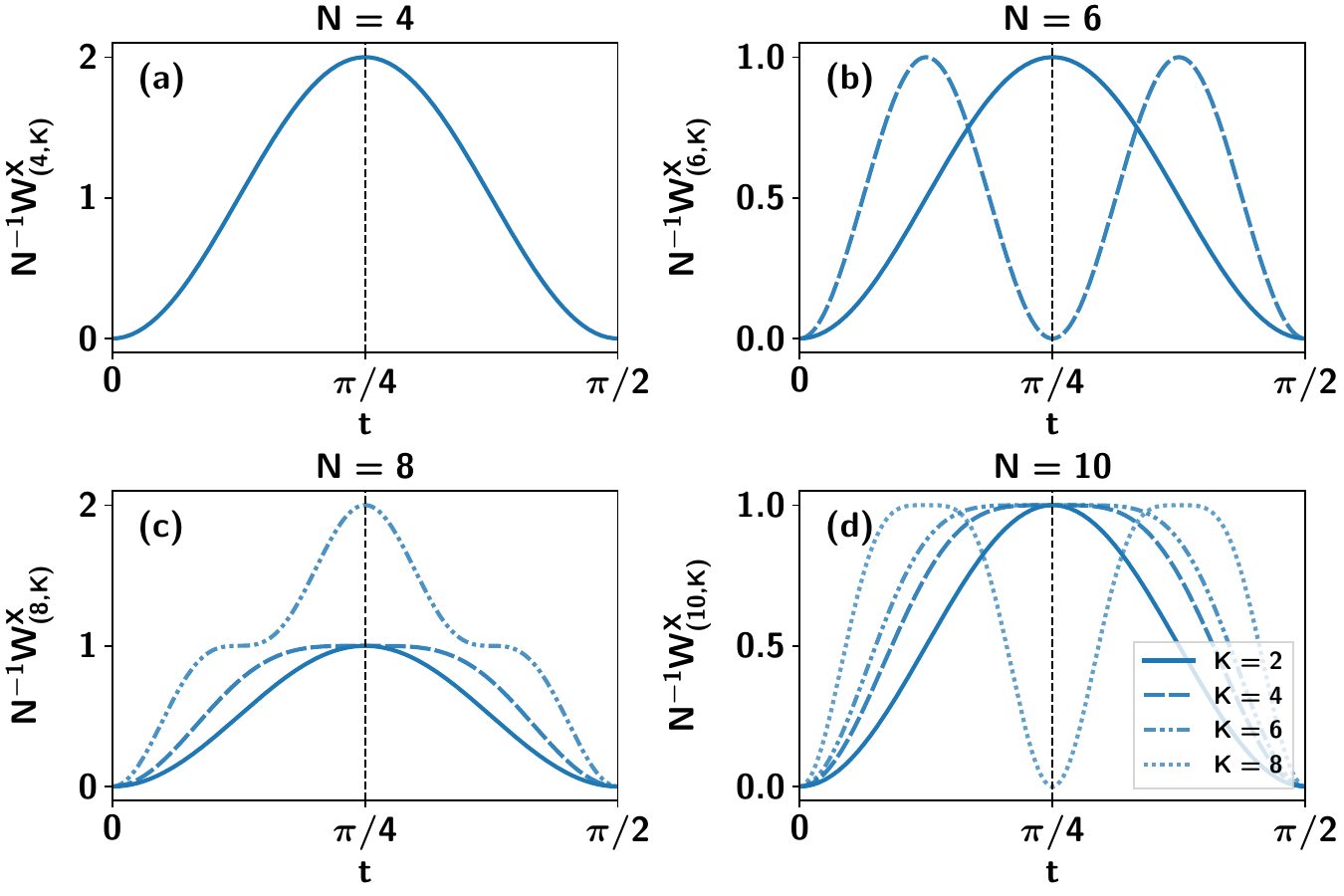}
    \caption{(Color online.) $N^{-1}W^X_{(N,K)}$ (ordinate) is plotted against $t$ (abscissa) for different \(K\)-regular charging Hamiltonian with (a) $N=4$, (b) $N=6$, (c) $N=8$, and (d) $N=10$. For a particular $N$, different line styles (solid, dashed, dashed-dotted and dotted) indicate different $K$ values. Note that for $K=K_{\max}$, $N^{-1}W^X_{(N,K)}$ shows different behavior than $K<K_{\max}$. At $t=\frac{\pi}{4}$, for $N=4$ and $8$ the maximum of $N^{-1}W^X_{(N,K)}$ is $2$ while for $N=6$ and $10$, it is $1$. All the axes are dimensionless.}
\label{fig:work_k_dependence_even_N}
\end{figure*}

\section{\texorpdfstring{$K$}{K}-regular charger on a local battery: Quantum advantage}
\label{sec:K_uniform_charger}

In this paper, we will be specifically interested in  \emph{local} battery Hamiltonians, 
  \begin{eqnarray}
      H_B^A =\sum_{i=1}^N A_i,
      \label{eqn:Bat_Hamil}
  \end{eqnarray}
where the battery corresponds to an ensemble of $N$ spin-$\frac{1}{2}$ particles subjected to local magnetic fields, such that $H_B^X$ can be identified to be \(H_{(N,0)}\), and $H_B^Y$ and $H_B^Z$ can be obtained from $H_{(N,0)}$ via local unitary transformations 
\begin{eqnarray}
    H^A_{B}=\left[\otimes_{i=1}^NU^{A}_i\right]H_{(N,0)}\left[\otimes_{i=1}^NU^{A}_i\right]^\dagger,
\end{eqnarray}
for $A=Y,Z$, where $U^Y_i=\exp\left[-\text{i}\pi Z/4\right]$ is a counter-clockwise $\pi/2$ rotation around the $z$ axis, and $U^Z_i=H_d$, the Hadamard operation. Since $K^\prime=0$ for all local battery Hamiltonians, we denote the corresponding work and ergotropy as $W^A_{(N,K)}$ and $\mathcal{E}^A_{(N,K)}$ respectively, where the superscript $A$ corresponds to the choice of $A$ used in $H_B$. Similarly, the maximum average power, and the instantaneous power, in this case, are denoted by $\overline{P}^A_{(N,K)}$ and $P^A_{(N,K)}$.  
 



We now focus on 
$K$-regular chargers with $K>0$, starting with $H_B^A$ as the battery ($A=X,Y$ and $Z$) and $H_C=H_{(N,2)}$ as the charger. For this, the following proposition holds for arbitrary \(K\) and \(N\) (see Appendix~\ref{app:2_reg-proof} for  the proof).

\PROP{2}  \emph{The maximum stored work due to a $2$-regular charger scales linearly with the system-size.}

\noindent  The work stored in the battery is given as
\begin{eqnarray}
    W^X_{(N,2)}&=&N(1-\cos^22t),\nonumber\\
    W^Y_{(N,2)}&=&N(1-\cos^32t),\nonumber\\
    W^Z_{(N,2)}&=&N(1-\cos 2t).
    \label{eq:work_2_graph}
\end{eqnarray}
\noindent In Fig.~\ref{fig:avg_work}, we plot the work stored in the battery where the charging is performed with \(2\)-regular graph for different system sizes. We find out that this exactly matches with Eq. (\ref{eq:work_2_graph}) and the maximum work can be obtained when the battery is made of \(Z\) and \(Y\) spins. In contrast to system size $N$, the role of $K$ depends on the choice of $A$ while defining $H_B^A$. For $A=Z$,  the next proposition holds (see Appendix~\ref{app:prop2_proof} for the proof).  

\PROP{3} \emph{For $H_B^A$ defined with $A=Z$, the work-output due to a $K$-regular charger varies linearly with $N$, and is independent of $K$.}

Similar analysis can also be performed using $A=X$ and $Y$ also, although the detailed calculation is cumbersome. Our analysis supported by numerical investigations reveals specific dependence of $W_{(N,K)}^A$ on $K$ when $A\in\{X,Y\}$, in contrast to the case of $A=Z$, where $W_{(N,K)}^A$ is $K$-independent (Proposition 2). This is  consolidated in the following proposition. 

\begin{figure*}
    \centering
    \includegraphics[width=0.9\linewidth]{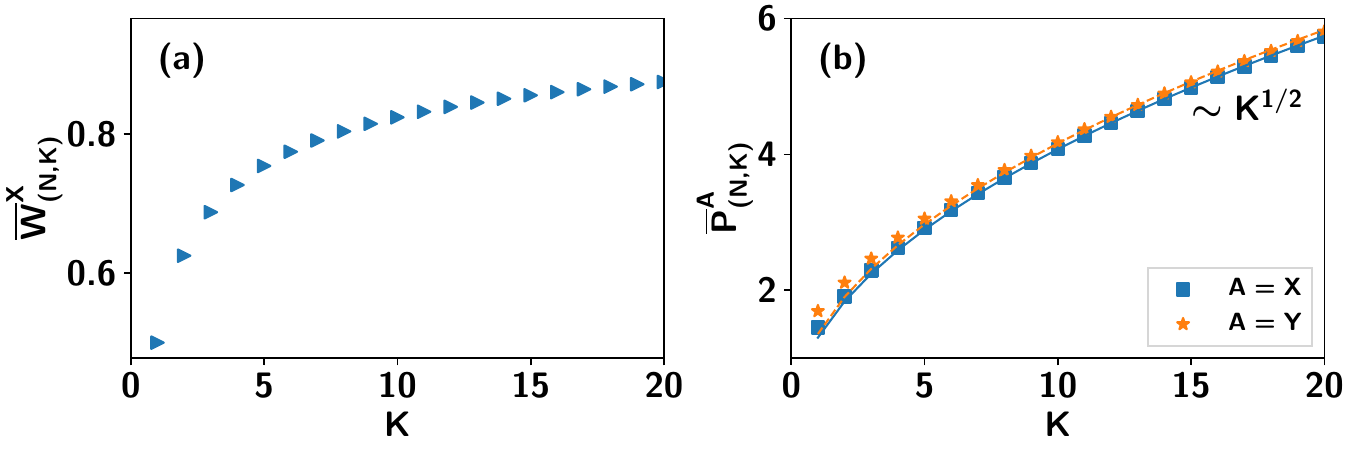}
    \caption{(Color online.) (a) Average work \(\overline{W}^X_{(N,K)}\) (y-axis) with respect to regularity, $K$ (x-axis). We find that   \(\overline{W}^X_{(N,K)}\) increases with regularity, \(K\) and saturates to a finite value, depicting the role of \(K\) in storing average work within a time domain. (b) Average power, $\overline{P}^A_{(N,K)}$ (y-axis)  against $K$ (x-axis) for the battery Hamiltonian $H_B^X$ and $H_B^Y$ with squares and stars respectively. For both the case $N=60$. All the axes are dimensionless.}  
    \label{fig:avg_k}
\end{figure*}

\PROP{4} \emph{For $H_B^A$ defined with $A\in\{X,Y,Z\}$, the work-output due to a $K$-regular charger varies linearly with $N$, and depends on $K$ as
\begin{eqnarray}
    W_{(N,K)}^A &=& N\left(1-\cos^{K+r^A} 2t\right),
    \label{eq:wxyz}
\end{eqnarray}
where $r^X=0$, $r^Y=1$, and $r^Z=1-K$.} 

\noindent The general proof for arbitrary interaction order $K$ is analytically involved and becomes increasingly cumbersome. In \textbf{Proposition~2}, we provide an exact derivation of the work stored in QB for the case $K=2$. For general $K$, we instead verify the proposed expression numerically.  For \(A=X\) and \(N=10\), the numerically computed work output is in excellent agreement with Eq.~(\ref{eq:wxyz}) for all values of \(K\) considered. This numerical agreement strongly supports the validity of Eq.~(\ref{eq:wxyz}) for arbitrary \(K\) as shown in Fig.~\ref{fig:work_k_dependence_even_N}(d).

Eq.~(\ref{eq:wxyz}) indicates the maximization of $W^Y_{(N,K)}$ ($W^X_{(N,K)}$) occurring at $t=(2\ell+1)\pi/2$,  ($t=(2\ell+1)\pi/4$), with $\ell=0,1,2,\cdots$. To check how fast $W^A_{(N,K)}$ ($A\in\{X,Y\}$) attains its maximum, note that Taylor expansion of $W_{(N,K)}^A$ around $t=\pi/2$ (i.e., $\ell=0$) results in  
\begin{eqnarray}
    W_{(N,K)}^Y&=& N\bigg[2-2(K+1)\Delta t^2+2(K+1)\left(K+\frac{2}{3}\right)\nonumber\\&&\nonumber\times\Delta t^4+\mathcal{O}(\Delta t^{6})\big\}\bigg],
\end{eqnarray} 
with $\Delta t=t-\pi/2$, which, in the limit $t\to \pi/2-t_{\epsilon}$, leads to  $W_{(N,K)}^Y\simeq 2$ and  $t_{\epsilon}=(K+2/3)^{-1/2}$. Similar analysis for $W_{(N,K)}^X$ in the vicinity of $t=\pi/4$ (i.e., $\ell=0$) provides 
\begin{eqnarray}
  \nonumber  W_{(N,K)}^X&=& N\bigg[1-2^{K}\Delta t^{K}+\frac{2K}{3}2^K\Delta t^{K+2}\\&&\nonumber+\mathcal{O}\left(\Delta t^{K+4}\right)\bigg],
\end{eqnarray}
with $\Delta t=t-\pi/4$, leading to $W_{(N,K)}^X\simeq1$ in the limit $t\to \pi/4-t_\epsilon$ where $t_\epsilon=(2K/3)^{-1/2}$. This implies a faster approach of $W_{(N,K)}^X$ in comparison to  $W_{(N,K)}^Y$, as shown by the $K$-dependence of $t_\epsilon$ $\forall K\geq2$.

Note that as $K$ increases, the QB is able to retain maximum charge for a longer duration  for all $N$ when $A\in\{X,Y\}$. When $N$ is even, the charging process is qualitatively different for the cases of $K=K_{\max}$ and $K<K_{\max}$, as demonstrated in Fig.~\ref{fig:work_k_dependence_even_N}. For $N=2N^\prime$ with odd $N^\prime$, the period of $W^A_{(N,K_{\max})}$ for $A\in\{X,Y\}$, as a function of $t$, is half of the same corresponding to  $W^A_{(N,K<K_{\max})}$, while in the case of even $N^\prime$, both periods are equal. This difference originates from the fact that $\langle\psi_B|F^A|\psi_B\rangle\neq0$ in the case of $K=K_{\max}$, in contrast to the cases of $K<K_{\max}$ for which $F^A|\psi_B\rangle=|\psi_B^\perp\rangle$, when $A\in\{X,Y\}$. 

The next corollaries follow directly from Proposition 4. 

\CORL{1} \emph{For $H_B^A$ defined with $A\in\{X,Y,Z\}$, $\overline{P}^A_{(N,K)}\sim N$ while $K$ is fixed and finite, implying an absence of quantum advantage.} 

\noindent Nevertheless, one can still observe the role of  \(K\) in the charging process (see Eq.~(\ref{eq:wxyz})), as follows.  

\CORL{2} \emph{For $H_B^A$ defined with $A\in\{X,Y,Z\}$, \(\overline{P}^A_{(N,K)}\propto\sqrt{K}\) for $A=X,Y$ with a fixed and finite $N$, while $\overline{P}^Z_{(N,K)}$ is independent of $K$.}

\noindent This enhancement of power highlights the advantage of increasing \(K\) while charging a local quantum battery with a $K$-regular charger (see Fig.~\ref{fig:avg_k}(b)). In what follows, we prove that a quantum advantage can be obtained when $K\to N$ for large $N,K$.

\subsection{Asymptotic quantum advantage in power}
\label{subsec:scalinganal}

We now discuss the presence of quantum advantage in the maximum average power and the instantaneous power at the $t\to0$ in the asymptotic limit $N,K\to\infty$, with $K\to N$. 

\subsubsection{In maximum average power}
\label{subsubsec:avgpower}

It has been established that the \emph{maximum average power} typically scales nonlinearly with the system-size, i.e., \(\overline{P}^A_{(N,K)}\propto N^\beta\), with a quantum advantage being observed when \(\beta>1\)~\cite{Bera2020QB,gyhm2022,Gyhm2024}. Our next theorem proves the existence of quantum advantage in maximum average power of the local battery charged with $K$-regular graphs as long as $A=X,Y$ when $K\to N$ and $K,N$ are large, while no quantum advantage is obtained for $A=Z$ in the same limit. 

\THEO{1} \emph{For $H_B^A$ defined with $A\in\{X,Y,Z\}$, for $N,K\to\infty$ at $t\to 0$,  $\overline{P}^{X,Y}_{(N,K)}\sim N^{3/2}$ if $K\sim N$ while $\overline{P}^Z_{(N,K)}\sim N$.}
\begin{proof}
We begin with the expression of the work-output for $A=X$, as given in Eq. (\ref{eq:wxyz}):
\begin{equation}
    W_{(N,K)}^{X}(t) = N (1 - \cos^K 2t).
\end{equation}
To find the optimal charging time $t^*$ that yields the maximum of $\overline{P}_{(N,K)}^{X}$, we maximize the average power with respect to time by setting
\begin{equation}
    \nonumber\frac{d}{dt} \left[ t^{-1} W_{(N,k)}^{X}(t) \right] = 0, 
\end{equation}
leading to
\begin{equation}
    t^{-1}(1 - \cos^K 2t) - 2K \sin(2t) \cos^{K-1}(2t) = 0.
    \label{eqn:maximization_condition}
\end{equation}
The maximum value of power shifts towards \(t=0\) limit with increasing \(K\). Hence, a series expansion of the LHS of the Eq. (\ref{eqn:maximization_condition}) can be carried out for a short operational time interval $t \to 0$ to find the solution $t=t^*$ of the Eq. (\ref{eqn:maximization_condition}) along with $K\to\infty$. 
So the Eq. (\ref{eqn:maximization_condition}) can be expressed as 
\begin{equation}
    \nonumber\sqrt{2K}\xi(\sqrt{2K}t)=0
\end{equation}
where, $ \xi(x)=\sum_{n=1}^\infty (-1)^n
    \frac
    {(2n-1)}{n!} x^{2n-1}$.
These lead to the optimal time,
\begin{equation}
t^*\simeq\frac{x^*}{\sqrt{2K}}\,,
\label{eqn:opt_time}
\end{equation}
for $x^*=\sqrt{-\mathcal{W}_{-1}(\frac{-1}{2\sqrt{e}})-\frac{1}{2}}$, where $\mathcal W(x)$ is the Lambert $W$ function.
Substituting optimal charging time $t^*$ in the expression of $\overline{P}_{(N,K)}^{X}$, we get the maximum power 
\begin{equation}
\overline{P}_{(N,K)}^{X}(t^*)\simeq \alpha N \sqrt{K}\,,
\end{equation}
where $\alpha=\frac{2\sqrt{2}x^*}{1+2{x^*}^2}$. For $K\sim N$, i.e., in the asymptotic limit $N \to \infty$,  
\begin{equation}
    \overline{P}_{(N,K)}^{X}(t^*) \propto N\sqrt{K} \sim N^{3/2},
    \label{eq:quantum_advantage}
\end{equation}
thereby establishing a collective quantum advantage. The proof is similar for $A=Y$. However, no such benefit can be obtained for \(H_B^Z\) as $W^A_{(N,K)}(t)$ is independent of \(K\) (see Eq. (\ref{eq:wxyz})). 
\end{proof}

\subsubsection{In maximum instantaneous power}
\label{subsubsec:inspower}

Next, we analyze the advantage in the instantaneous power, as given in Theorem 2.  

\THEO{2} \emph{For $H_B^A$ defined with $A\in\{X,Y,Z\}$, for $N,K\to\infty$ at $t\to 0$,  $P^{X,Y}_{(N,K)}\sim N^{3/2}$ if $K\sim N$ while $P^Z_{(N,K)}\sim N$.}

\begin{proof}
We begin by evaluating the instantaneous power $P^A_{(N,K)}$ for $A=X$ from $W_{(N,k)}^{X}(t)$ as 
\begin{eqnarray}
    P^A_{(N,K)} &=& \frac{d}{dt} \left[ N(1 - \cos^K 2t) \right],\nonumber\\ &=& 2N K \sin(2t) \cos^{K-1}(2t).
    \label{eqn:insp}
\end{eqnarray}
For $A=X$, the instantaneous power reads
\begin{eqnarray}
 \nonumber P(t)&=&2NK\cos^{K-1}(2t) \sin(2t)\\
 &=& 2NK\sin(2t)e^{(K-1)\ln\cos(2t)}  
\end{eqnarray}
If $K t^2\leq c$ with \(c\) being a finite positive constant, and in that case when $K\to\infty$ we obtain
\begin{equation}
P(t)\simeq 4 NK te^{-2Kt^2},
\label{eqn:inspower}
\end{equation}
since the neglected terms go to zero. Then, we get the derivative of the Eq. (\ref{eqn:inspower}) as \(P'(t)\simeq 4 NK (1-4Kt^2)e^{-2Kt^2}\). For the optimal charging time $t=t^*$, $P'(t^*)=0$. This leads to
\begin{equation}
t^*\simeq \frac{1}{2\sqrt{K}}
\label{eqn:opt_ins}
\end{equation}
which satisfies $Kt^2\leq c$. Substituting Eq. (\ref{eqn:opt_ins}) into Eq. (\ref{eqn:insp}), we get the maximum instantaneous power
\begin{equation}
\nonumber P^{max}= P^X_{(N,K)}(t^*)\simeq \frac{2}{\sqrt{e}}N\sqrt{K},
\end{equation}
which implies 
\begin{equation}
    P^X_{(N,K)}(t^*)\propto  N^{3/2}
\end{equation}
for $K\sim N$, i.e, at $N,K\to\infty$. Similar proof stands for $A=Y$, while $P^Z_{(N,K)}\propto N$ follows from the expression for $W^Z_{(N,K)}(t)$. 
\end{proof}
\noindent We note that the maximum instantaneous power is larger than the maximum average power since $2/\sqrt{e}>\alpha$, as expected.

\noindent\textbf{Note 1.} It is worthwhile to note that the Corollary 1 and 2 remain valid in the cases of fixed $K$ ($N$) with $N\to\infty$ ($K\to\infty$) respectively, as is clear from the proofs of Theorems 1 and 2.  

\noindent\textbf{Note 2.} Following Proposition 1, the quantum advantage, reported in Theorem 1 and 2 can also be obtained via charging locally while the initial state is the ground state of an interactive \(K\)-regular graph Hamiltonian.

\subsection{System size-agnostic average work} 

We now define the average work integrated over a complete period $T^A$ as
\begin{eqnarray}
    \overline{W}_{(N,K)}^A = \frac{1}{N T^A} \int_0^{T^A} W_{(N,K)}^A dt,
    \label{eq:avg_work}
\end{eqnarray}
where $T^A$ depends on the choice of $A$, and is given by 
\begin{eqnarray}
    T^X&=&\pi/2\quad \text{and} \quad T^Y=T^Z= \pi.
\end{eqnarray}
The quantity \(\overline{W}_{(N,K)}^A\) denotes the average work stored in the battery over a given time interval. We say that a quantum advantage in average work is achieved when
\(\overline{W}_{(N,K)}^A>1\).
Substituting Eq. (\ref{eq:wxyz}) in Eq. (\ref{eq:avg_work}), one obtains

\begin{eqnarray}
    \overline{W}_{(N,K)}^X =1 - \frac{\Gamma(K + 1/2)}{\sqrt{\pi}\, \Gamma(K+1)},\;\; \overline{W}_{(N,K)}^Y =\overline{W}_{(N,K)}^Z=1,\nonumber\\  
\end{eqnarray}
where $\Gamma(.)$ is the Gamma function.  Note that $\overline{W}^A_{(N,K)}$ is maximized at $K = K_{\max}$ for a given $N$.

\noindent\textbf{Remark.} While it is logical to investigate the charging of a $K^\prime$-regular battery with a $K$-regular charger in the wake of the performance of the $K$-regular charger with a local battery, the dependence of stored work as well as extractable energy on $K^\prime$ and $K$ is complex, and obtaining analytical closed forms for $W_{(N,K^\prime,K)}$ is difficult.  While our initial investigation with $N\leq 10$ indicates that $W_{(N,K^\prime,K)}$ remains unchanged as long $\Delta = |K-K^{\prime}|$ is constant for $N\neq 4N^\prime$ with $N^\prime=1,2,\cdots$, no such trend is observed in the case of  $N\neq 4N^\prime$. However, concrete conclusion requires extensive numerical analysis with large $N$, which is challenging.

\begin{figure*}
\includegraphics[width=\linewidth]{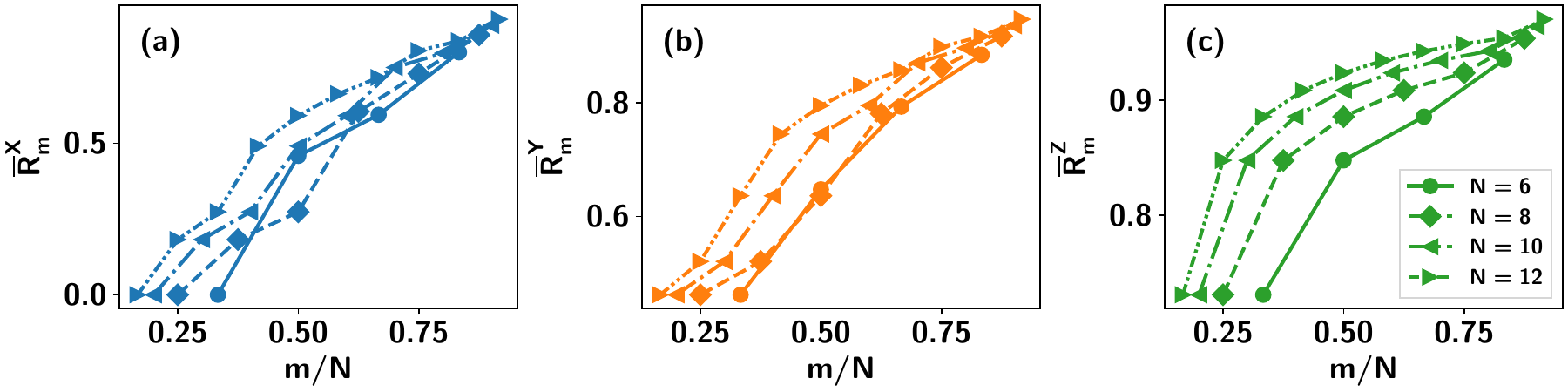}
\caption{(Color online.) Fraction of the extractable energy, $\overline{R}^A_m$ (vertical axis) with the ratio, $m/N$ (horizontal axis) for the regularity $K=2$. The subplots are for the battery Hamiltonian with (a) $A=X$, (b) $A=Y$, and (c) $A=Z$. For a particular battery Hamiltonian $H_B^A$, different line-points indicates different $N$ ranging from $N=6  \, (\text{circles}),8\,   (\text{squares}),10$ (left-arrowed triangles) and $12$ (right-arrowed triangles). Note that $\overline{R}^Z_m$ is independent of $N$ differ from $A=X$ and $Y$ where there is a dependency of $N$. All the axes are dimensionless.}
\label{fig:extract_work}
\end{figure*}

\section{Fraction of extractable energy}
\label{sec:fracen}
 
In a many-body quantum battery, it is often difficult to control the entire system of size \(N\), while  manipulating only a subset of \(m\) parties \(m<N\) may be feasible. Further, since the state of the accessible part of the battery is mixed, the work stored in these \(m\) parties generally differs from their ergotropy.  In this scenario,  we define the ratio (cf. \cite{rossini_prb_2019})
\begin{equation}
\overline{R}_m^A=\frac{\int_0^{T}\mathcal{E}^A_m(t)\,dt}{\int_0^{T}W^A_{(m,K)}(t)\,dt},
\label{eqn:extract}
\end{equation}
quantifying the fraction of the extractable energy (i.e., \emph{ergotropy}) from the \(m\) accessible parties to the amount of energy stored in the entire system. Here, we denote the local battery Hamiltonian of the accessible subsystem as
\(H_B^A(m)=\sum_{i=1}^m A_i\), and \(\rho_m(t)\) is the reduced density matrix of the \(m\)-party subsystem, obtained from \(\rho(t)\) by tracing out the inaccessible \(N-m\) sites in Eqs.~(\ref{eq:maximum_extractable_work}) and (\ref{eq:ergo}). Note that for a fully accessible battery \((m=N)\), we always obtain \(\overline{R}_m^A=1\), regardless of the choice of the averaging interval, $T$.

We analyze the scaling of \(\overline{R}_m^A\) with respect to \(m\) and \(K\), by fixing the charging Hamiltonian as \(H_C=H_{(N,K)}\) for different $N$  (\(N\leq12\)). For \(T=\pi\), \(R_m^A\) increases with \(m/N\), and saturates to a finite constant (see Fig.~\ref{fig:extract_work}) in the limit $m/N\to 1$.
We observe that \(\overline{R}_m^Z>\max\{\overline{R}_m^X,\overline{R}_m^Y\}\) even for \(m/N<1/2\), indicating that employing \(H_B^Z\) as the battery Hamiltonian allows greater energy extraction from the subsystem, as illustrated in Fig.~\ref{fig:extract_work}(c). To understand this, note that the battery $H_B^Z$ is initially prepared in a fully down-polarized state, denoted by \(\ket{0}^{\otimes N}\). As the system evolves, correlations are generated among the spins, which contribute to the ergotropy. Simultaneously, the charging Hamiltonian $H_{(N,K)}$ flips the spins upward, thereby increasing the extractable energy. The time evolution of the full system is given by
\begin{eqnarray}
    \ket{\Psi(t)} &=& \text{e}^{-\text{i} H_{(N,K)} t}\ket{0}^{\otimes N}, \nonumber \\
    &=& \bigotimes_{i=1}^{N}\left(\cos t\,\mathbb{I}-\mathrm{i}\sin t H_{(i,K)}\right)\ket{0}^{\otimes N},
\end{eqnarray}
where \(\mathbb{I}\) is the identity operator of dimension \(2^3\).  
At \(t=t_n=(2n+1)\pi/2\), the evolved state simplifies to
\begin{equation}
    \ket{\Psi(t_n)} = (-\mathrm{i})^N \ket{1}^{\otimes N},
\end{equation}
since the operator \(\bigotimes_{i=1}^{N} H_{(i,K)}\) can be expressed as \(\bigotimes_{i=1}^{N} H_{(i,K)} = \bigotimes_{i=1}^{N} X_i\), which flips all spins up, implying that \(\ket{\Psi(t_n)}\) contains no spin-spin correlations, and leads to pure \(m\)-party reduced states.  Therefore, the ergotropy of the \(m\)-party subsystem coincides with the energy stored in that subsystem at \(t=t_n\), i.e.,
\begin{equation}
    \mathcal{E}_m^Z(t_n)=W^Z_{(m,K)}(t_n).
\end{equation}
Such a correspondence does not occur for \(A=X\) or \(=Y\), where residual correlations persist in the reduced states. 
Further, in contrast to $\overline{R}_m^X$ and $\overline{R}_m^Y$,   $\overline{R}_m^Z$ is independent $N$.

\section{Collective \texorpdfstring{$K$}{K}-regular chargers on a local battery}
\label{sec:collective}

\begin{figure}
   \centering
   \includegraphics[width=\linewidth]{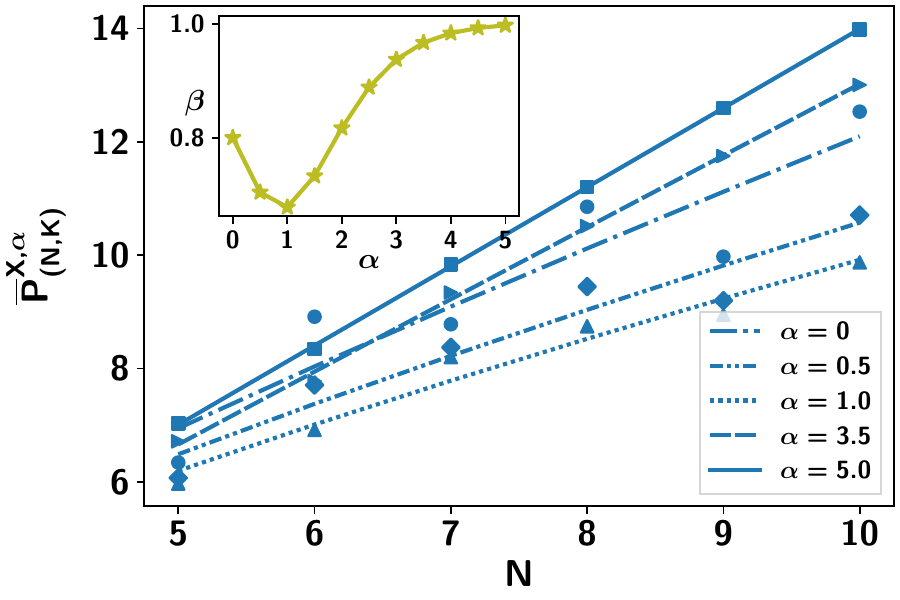}
   \caption{(Color online.) Average power, $\overline{P}^{X,\alpha}_{(N,K)}$ (y-axis) against system-size, $N$ (x-axis) for the battery Hamiltonian $H_B^X$ with the different line-style indicate different values of fall rate $\alpha$. $\overline{P}^{X,\alpha}_{(N,K)}$ scales with $N^\beta$ where $\alpha<5$ indicates sublinear scaling, i.e., $\beta<1$. In the inset, we plot   $\beta$ (y-axis) with the fall rate, $\alpha$. $\beta$ saturates to unity  for $\alpha>5$ which implies no quantum advantage for $\alpha<5$. Interestingly, $\alpha=1$ has the lowest value of $\beta$. All the axes are dimensionless. }
    \label{fig:weighted}
\end{figure}

In this section, we address the following question- \emph{''Does going beyond the set of stabilizer Hamiltonians $H_{(N,K)}$ provide an advantage in charging local batteries?"} To investigate this, motivated by Eq.~(\ref{eq:non_commutation}), we consider a Hamiltonian constituted  of the $H_{(N,K)}$ corresponding to all possible $K$-regular graphs with fixed $N$ as the charging Hamiltonian. It is given by  

\begin{eqnarray}
    H_N^\alpha&=&\sum_{K=2}^{K_{\max}}J_K H_{(N,K)},
    \label{eq:charger_K_collective}
\end{eqnarray}
where the weight $J_K$ can be interpreted as the strength of the interaction corresponding to $H_{(N,K)}$ for a fixed $K$. In this paper, we assume a decreasing strength of interaction $J_K$ with increasing $K$ as $J_K=\left[\sum_{K=2}^{K_{\max}}(K/2)^{-\alpha}\right]^{-1}(K/2)^{-\alpha}$, where $\alpha$ ($>0$) is the fall rate. Note that $\alpha=0$ corresponds to the case of equal and unit weight corresponding to all $H_{(N,K)}$, leading to $H_N^0=\sum_{K=2}^{K_{\max}}H_{(N,K)}$. For the charging Hamiltonian $H_N^\alpha$, we plot $\overline{P}_{(N,K)}^{A,\alpha}$ (Eq.~(\ref{eqn:avg_power}))  as a function of $N$ with different values of $\alpha$ (see Fig.~\ref{fig:weighted}). For $A=X$, our analysis shows that $\overline{P}_{(N,K)}^{X,\alpha}$ scales with $N^{\beta}$. While $\beta<1$ for small $\alpha$, it increases with $\alpha$  and saturates to unity for $\alpha>5$, demonstrating the lack of super-extensive scaling of $\overline{P}_{(N,K)}^{X,\alpha}$ with $N$ even when the stabilizer nature of the charging Hamiltonian is removed.

\section{Conclusion}
\label{sec:conclusion}

Summarizing, our work systematically explored the potential of a class of graphs, namely, the $K$-regular graphs, to obtain quantum advantage  within the field of quantum thermodynamics, specifically in designing quantum batteries. We introduced a general framework for designing quantum chargers built using generators of $K$-regular graph, while the battery consists of an ensemble of spins initially aligned along a chosen direction. Within this setting, we explored how the pattern of connections between vertices influences the performance of the energy-storing devices. Beyond the extractable work stored in the battery, we proved that, in the thermodynamic limit, both the instantaneous power and the maximum average charging power exhibit superlinear scaling, provided the graph connectivity increases with the system-size. These results identify graph-state chargers as another experimentally implementable quantum battery architecture that can exhibit a genuine quantum advantage in charging performance. We also studied scenarios in which only a portion of the battery is accessible,  and evaluated the fraction of  energy extractable from the subsystem. Our analysis revealed that the local extractable work increases monotonically with the size of the accessible region, showing that even partial access to a graph-structured battery retains its usefulness and scales consistently with the subsystem-size. When the battery Hamiltonian is oriented along the \(z\)-direction, this fraction remains independent of system-size; in contrast, a pronounced dependence on system-size  emerges, when the initial battery state is prepared along other complementary directions. Going beyond idealized uniform graphs, we incorporated a collection of stabilizer graph generators corresponding to different $K$-regular graphs as charger with $K$-dependent interaction strengths governed by power-law decay, and found that the average power scales linearly with the system-size. 

\acknowledgments

The authors acknowledge the cluster computing facility at Harish-Chandra Research Institute and the use of \href{https://github.com/titaschanda/QIClib}{QIClib} -- a modern C++ library for general purpose quantum information processing and quantum computing. This research was supported in part by the ``INFOSYS scholarship for senior students''. A.K.P and A.S.D acknowledge the support from the Anusandhan National Research Foundation (ANRF) of the Department of Science and Technology (DST), India, through the Core Research Grant (CRG) (File No. CRG/2023/001217, Sanction Date 16 May 2024). A.S.D. acknowledges support from the project entitled "Technology Vertical - Quantum Communication'' under the National Quantum Mission of the Department of Science and Technology (DST)  (Sanction Order No. DST/QTC/NQM/QComm/2024/2 (G)).

\appendix

\section{Proof of Proposition 1}
\label{app:proof_prop_1}
Consider two stabilizer Hamiltonians \(H_{(N,K^\prime)}\) and \(H_{(N,K)}\) with \(K\neq K^\prime\), where the former (latter) acts as the battery (charger), having the ground states \(\ket{\Psi_{K^\prime}}\) and \(\ket{\Psi_K}\) respectively, satisfying (see Eq.~(\ref{eq:stab-to-stab-unitary}))
\begin{equation}
    \ket{\Psi_{K^\prime}}=\mathcal{U}_{(K^\prime,K)}\ket{\Psi_K}.
    \label{eq:ground_state_connection}
\end{equation}
Since all $H_{(N,K)}$ share the same eigenvalues, their ground state energies are identical, hence, the initial energy is given as \(\text{Tr}[\rho(0)H_B]=-N\).  Therefore, initializing the battery in its ground state, it suffices to analyze the time evolution to determine the stored energy, which is given by 
\begin{widetext}
\begin{eqnarray}
    \bra{\Psi_{K^\prime}}\text{e}^{\text{i}H_{(N,K)}t}H_{(N,K^\prime)}\text{e}^{-\text{i}H_{(N,K)}t}\ket{\Psi_{K^\prime}}
    &=& \bra{\Psi_K}\mathcal{U}^\dagger_{(K^\prime,K)} \text{e}^{\text{i}H_{(N,K)}t}\mathcal{U}_{(K^\prime,K)}H_{(N,K)}\mathcal{U}^\dagger_{(K^\prime,K)} \text{e}^{-\text{i}H_{(N,K)}t}\mathcal{U}_{(K^\prime,K)}\ket{\Psi_K}, \nonumber\\
    &=& \bra{\Psi_K} \text{e}^{\text{i}\mathcal{U}^\dagger_{(K^\prime,K)} H_{(N,K)}\mathcal{U}_{(K^\prime,K)}t}H_{(N,K)} \text{e}^{-\text{i}\mathcal{U}^\dagger_{(K^\prime,K)} H_{(N,K)}\mathcal{U}_{(K^\prime,K)}t}\ket{\Psi_K}, \nonumber\\
    &=& \bra{\Psi_K} \text{e}^{\text{i}H_{(N,K^\prime)}t}H_{(N,K)} \text{e}^{-\text{i}H_{(N,K^\prime)}t}\ket{\Psi_K},\nonumber\\
\end{eqnarray} 
\end{widetext}
where we have used Eq.~(\ref{eq:ground_state_connection}) and  \(\mathcal{U}_{(K^\prime,K)}=\mathcal{U}^\dagger_{(K^\prime,K)}\),  leading to \(W_{(N,K^\prime,K)}=W_{(N,K,K^\prime)}\). Hence the proof. 

\section{Proof of Proposition 2}
\label{app:2_reg-proof}

The maximum work stored in the battery at any arbitrary time $t$ (see Eq.~(\ref{eq:maximum_extractable_work})) is given by
  \begin{eqnarray}
   W^A_{(N,K)}&=& \langle\psi_B|U^\dagger H_B^A U \ket{\psi_B}-\bra{\psi_B} H_B^A \ket{\psi_B},\nonumber\\
   &=&\langle\psi_B|U^\dagger H_B^A U \ket{\psi_B}+N.
   \label{eqn: work}
\end{eqnarray}

Note that
\begin{eqnarray}
    U^\dagger H_B^A U&=&\sum_{k=1}^N U^\dagger A_k U=\sum_{k=1}^N\mathcal{A}_k, 
\end{eqnarray}
where we have defined $\mathcal{A}=U^\dagger A U$, such that $\mathcal{A}\in\{\mathcal{X},\mathcal{Y},\mathcal{Z}\}$ while $A\in\{X,Y,Z\}$, with $\mathcal{X}=U^\dagger X U$, and similarly for $Y$ and $Z$. Let us first consider $A=Z$, which, using $\left[Z_k,\exp\left\{-\text{i}H_{(j,2)}t\right\}\right]=2t \sin t Z_{k-1}Y_kZ_{k+1}\delta_{j,k}$, $\delta_{j,k}$ being the Kronecker's delta, becomes 
\begin{eqnarray}
    \mathcal{Z}_k&=&U^\dagger Z_k U,\nonumber\\
    &=& \cos 2t Z_k+2\sin 2t Z_{k-1}Y_kZ_{k+1},\nonumber\\
    &=&\cos 2t Z_k+F^Z,
\end{eqnarray}
where we have defined $F^Z=2\sin 2t Z_{k-1}Y_kZ_{k+1}$. Similar calculations with $A=X$ and $Y$ lead respectively to 
\begin{eqnarray}
    \mathcal{X}_k&=&\cos^2 2t X_k+F^X,\nonumber\\
    \mathcal{Y}_k&=&\cos^3 2t Y_k+F^Y,
\end{eqnarray}
where
\begin{eqnarray}
    F^X&=&-\frac{1}{4}\sin 4t Z_{k-2}X_{k-1}Y_k\nonumber\\
    &&-\sin 2t \cos^2 t Y_{k}X_{k+1}Z_{k+2}\nonumber\\
    &&+\left(\sin^2 t+\sin2 t\right)Z_{k-2}X_{k-1}Y_{k}X_{k+1}Z_{k+2}\nonumber\\
    &&-\sin^2 2tZ_{k-2}X_{k-1}X_{k}X_{k+1}Z_{k+2},\\
    F^Y&=& \sin2t\cos^22t\big(Z_{k-2}X_{k-1}X_k\nonumber\\&&-Z_{k-1}Z_kZ_{k+1}+X_kX_{k+1}Z_{k+2}\big)\nonumber\\
    &&-\sin^22t\cos2t\big(Z_{k-2}Y_{k-1}Z_{k+1}\nonumber\\&&+Z_{k-1}Y_{k+1}Z_{k+2}+Z_{k-2}X_{k-1}Y_kX_{k+1}Z_{k+2}\big)\nonumber\\
    &&-\sin^32tZ_{k-2}Y_{k-1}Z_kY_{k+1}Z_{k+2}.
\end{eqnarray}

For $N=3$ and $4$, explicit calculations using the above expressions lead to
\begin{eqnarray}
    W^Y_{(3,2)}&=&6\sin^2 3t\nonumber\\
    W^X_{(4,2)}&=&4\left(1+\cos^2 2t\right).
\end{eqnarray}
In all other cases,  $F^A$ projects the ground state into its orthogonal eigenstate $\ket{\psi_B^{\perp}}$ of $H_B^A$ (i.e., $F^A\ket{\psi_B}=\ket{\psi_B^{\perp}}$), and, therefore, does not contribute to the extractable work, leading to 
\begin{eqnarray}
    W^X_{(N,2)}&=&N(1-\cos^22t),\nonumber\\
    W^Y_{(N,2)}&=&N(1-\cos^32t),\nonumber\\
    W^Z_{(N,2)}&=&N(1-\cos 2t).
\end{eqnarray}
Hence the proof. 

\section{Proof of Proposition 3}
\label{app:prop2_proof}

The proof exploits the Baker–Campbell–Hausdorff (BCH) formula, given for $U=\exp\left[-\text{i}H_{(N,K)}t\right]$ by 
\begin{eqnarray}
\text{e}^{\text{i}H_{(N,K)}t} H_B^A \text{e}^{-\text{i}H_{(N,K)}t}&=&
  H_B^A+ \left[\text{i}tH_{(N,K)},H_B^A\right]\nonumber\\ &&+\frac{1}{2}\left[\text{i}tH_{(N,K)},\left[\text{i}tH_{(N,K)},H_B^A\right]\right]\nonumber\\&& +\ldots\nonumber\\&=&\sum_{n=0}^\infty\frac{1}{n!}\left[\text{i}tH_{(N,K)},H_B^A\right]_n,
  \label{eqn:BCH}
\end{eqnarray}
where the operator $[\text{i}tH_{(N,K)},H_B^A]_n$ is defined as \small 
\begin{eqnarray}
    \left[\text{i}tH_{(N,K)},H_B^A\right]_n=\left[\text{i}tH_{(N,K)},\left[\text{i}tH_{(N,K)},H_B^A\right]\right]_{n-1},
\end{eqnarray}\normalsize 
with
\begin{eqnarray}
    \left[\text{i}tH_{(N,K)},H_B^A\right]_1=\left[\text{i}tH_{(N,K)},H_B^A\right],
\end{eqnarray}
and 
\begin{eqnarray}
    \left[\text{i}tH_{(N,K)},H_B^A\right]_0=H_B^A.
\end{eqnarray}
Here, the subscript, $n$,  denotes the iteration number. For $A=Z$, we get
\begin{eqnarray}
    \left[\text{i}tH_{(N,K)},H_B^Z\right]_{2n+r}=(2t)^n\left[\text{i}tH_{(N,K)},H_B^Z\right]_{r},
\end{eqnarray}
for $r\in\{0,1\}$. Using these, Eq. (\ref{eqn:BCH}) simplifies to
\begin{eqnarray}
    U^\dagger H_B^Z U &=& H_B^Z\cos 2t  +\frac{\left[\text{i}H_{(N,K)},H_B^Z\right]t}{2t}\sin 2t,  
\end{eqnarray}
leading to
\begin{eqnarray}
    W^Z_{(N,K)}=2N\sin^2 t,
\end{eqnarray}
since $\left\langle\psi_B\right|\left[\text{i}tH_{(N,K)},H_B^Z\right]\left|\psi_B\right\rangle=0$. Hence the proof.

\bibliography{ref}

\end{document}